# Kaluza-Klein Theory as a Dynamics in a Dual Geometry

**Avi Gershon[1] and Lawrence Horwitz [1, 2, 3]**
[1] School of Physics, Tel Aviv University, Ramat Aviv 68878, Israel
[2] Department of Physics, College of Judea and Samaria, Ariel Israel
[3] Department of Physics, Bar Ilan University, Ramat Gan, Israel

*Abstract*

It has been shown that the orbits of motion for a wide class of non-relativistic Hamiltonian systems can be described as geodesic flow on a manifold and an associated dual. This method can be applied to a four dimensional manifold of orbits in space-time associated with a relativistic system. One can study the consequences on the geometry of the introduction of electromagnetic interaction. We find that resulting geometrical structure in the dual space is that of Kaluza and Klein.

**1. Introduction**

We start with a brief review of the geometry associated with a non-relativistic Hamiltonian system. Gutzwiller [1, 2] has pointed out that the Hamiltonian equations for a 3D Hamiltonian system of the form (we use the summation convention).

$$H = \frac{1}{2m} g_{ij} p^i p^j, \tag{1}$$

where $g_{ij}$ is a function of x alone, which is assumed to have an inverse $g^{ij}$, results in orbits described by the geodesic equation

$$\ddot{x}_i = -\Gamma_i^{jk} \dot{x}_j \dot{x}_k, \tag{2}$$

where what can be understood as a connection form is given by

$$\Gamma_i^{jk} = \frac{1}{2} g_{il} \left( \frac{\partial g^{lj}}{\partial x_k} + \frac{\partial g^{lk}}{\partial x_j} - \frac{\partial g^{jk}}{\partial x_l} \right). \tag{3}$$

The Hamilton equations therefore imply equations of motion that are of the form of a geodesic with a connection form that is compatible with this metric, and therefore we have the structure of a geometry covariant under local diffeomorphisms [2]. Thus, we have a space with Riemannian curvature induced by its dynamical geodesic structure.

It has been shown [1], that a Hamiltonian of the form

$$H = \frac{1}{2m} (p^i)^2 + V(x) \tag{4}$$

can be put into the form of (1) by defining, in, a conformal metric (see [3] for other methods of introducing geometrical structures),

$$g_{ij} = \frac{E}{E - V(x)} \delta_{ij}. \tag{5}$$

The equations of motion **(2)**, in what we shall consider as the space locally dual to the original Hamiltonian coordinates, do not, however, coincide with the Hamiltonian equations generated by **(4)**. If we introduce the local mapping

$$\dot{x}^i = g^{ij}\dot{x}_j \tag{6}$$

on the tangent space, one obtains a geodesic type equation for the $\{\dot{x}^j\}$. For the choice of coordinates for which **(5)** is valid, one obtains the Hamilton equations implied directly by **(4)**. The geodesic deviation obtained from the geodesic relation for $\{\dot{x}^j\}$ were found to provide an effective criterion for the stability of the motion determined by the Hamiltonian **(4)**.

In the following, we study the analogous mapping of a relativistic Hamiltonian with electromagnetic interaction to a Gutzwiller type structure. We shall then show that the resulting structure is that of Kaluza and Klein. Before dealing with the full relativistic structure, we first study the analogous non-relativistic problem.

## 2. The general method with gauge field in 3d Riemannian geometry.

### 2.1 Geometrical form of the 3D electromagnetic equations

Consider the general case of a Hamiltonian of the form

$$H = \frac{1}{2m}\eta_{ij}(p^i - eA^i)(p^j - eA^j) - eA^0, \tag{7}$$

where $p^\mu$ is momentum, $e$ is the elementary charge, $A^\nu(x,\tau)$ is the usual Maxwell field and $\eta_{ij}$ is Kronecker $\delta_{ij}$. With the substitution

$$g_{ij} = \frac{E}{E + eA^0}\eta_{ij}, \tag{8}$$

one obtains

$$H = \frac{1}{2m}g_{ij}(p^i - eA^i)(p^j - eA^j), \tag{9}$$

The resulting geodesic equation of motion is:

$$\ddot{x}_j = -\Gamma^{lm}_j \dot{x}_l \dot{x}_m - g_{jk}\left(\frac{\partial g^{lk}}{\partial t} + \frac{e}{m}\frac{\partial A^k}{\partial x_l} - \frac{e}{m}\frac{\partial A^l}{\partial x_k}\right)\dot{x}_l - g_{jk}\frac{e}{m}\frac{\partial A^k}{\partial t} \tag{10}$$

### 2.2 Curvature from the geodesic deviation

As a method of identifying the extra terms in **(10)**, we obtain the curvature tensor on this manifold by computing the geodesic deviation. To do this, let us look at a small deviation between two close-by orbits $y(\tau), x(\tau)$ with

$$\xi^i = y^i - x^i. \tag{11}$$

Since by **(10)**,

$$\ddot{y}_i = -\Gamma(x+\xi)_i^{jk}(\dot{x}+\dot{\xi})_k(\dot{x}+\dot{\xi})_j + \frac{e}{m}g(x+\xi)_{bi}f(x+\xi)^{bk}(\dot{x}+\dot{\xi})_k +$$

$$+g(x+\xi)_{bi}\left(-\frac{e}{m}\frac{\partial A(x+\xi)^b}{\partial t} - \frac{\partial g(x+\xi)^{bj}}{\partial t}(\dot{x}+\dot{\xi})_j\right). \tag{12}$$

By retaining only first order terms in $\xi$, we obtain

$$\ddot{\xi}_i = -\Gamma_i^{jk}(\dot{x}_j\dot{\xi}_k + \dot{\xi}_j\dot{x}_k) - \frac{\partial \Gamma_i^{jk}}{\partial x_l}\xi_l\dot{x}_j\dot{x}_k +$$

$$+\left(\frac{\partial}{\partial x_\nu}\left(\frac{e}{m}g_{bi}F^{bj}\right) - \frac{\partial}{\partial x_k}\left(g_{bi}\frac{\partial g^{bj}}{\partial j}\right)\right)\xi_k\dot{x}_j + \tag{13}$$

$$+\left(\frac{e}{m}g_{bi}F^{bj} - g_{bi}\frac{\partial g^{bj}}{\partial t}\right)\dot{\xi}_j - \frac{\partial}{\partial x_j}\left(\frac{e}{m}g_{bi}\frac{\partial A^b}{\partial t}\right)\xi_j.$$

The covariant derivative of the geodesic deviation $\xi$, is given by

$$\frac{D\xi_i}{D\tau} = \frac{d\xi_i}{d\tau} + \Gamma_i^{jk}\frac{\partial x_j}{\partial t}\xi_k = \dot{\xi}_i + \Gamma_i^{jk}\dot{x}_j\xi_k. \tag{14}$$

The second covariant derivative is then

$$\frac{D^2\xi_i}{Dt^2} = \frac{D(\dot{\xi}_i + \Gamma_i^{jk}\dot{x}_j\xi_k)}{Dt} = \frac{d(\dot{\xi}_i + \Gamma_i^{jk}\dot{x}_j\xi_k)}{dt} + \Gamma_i^{lb}\dot{x}_l(\dot{\xi}_b + \Gamma_b^{jk}\dot{x}_j\xi_k) =$$

$$= \ddot{\xi}_i + \Gamma_i^{jk}\ddot{x}_j\xi_k + \Gamma_i^{jk}\dot{x}_j\dot{\xi}_k + \frac{\partial \Gamma_i^{jk}}{\partial t}\dot{x}_j\xi_k \tag{15}$$

$$+ \frac{\partial \Gamma_i^{jk}}{\partial x_b}\dot{x}_b\dot{x}_j\xi_k + \Gamma_i^{jk}\dot{x}_j\dot{\xi}_k + \Gamma_i^{lb}\dot{x}_l\Gamma_b^{jk}\dot{x}_j\xi_k,$$

i.e.,

$$\ddot{\xi}_i = \frac{D^2\xi_i}{Dt^2} - \Gamma_i^{jk}\ddot{x}_j\xi_k - \Gamma_i^{jk}\dot{x}_j\dot{\xi}_k - \frac{\partial \Gamma_i^{jk}}{\partial t}\dot{x}_j\xi_k -$$

$$-\frac{\partial \Gamma_i^{jk}}{\partial x_b}\dot{x}_b\dot{x}_j\xi_k - \Gamma_i^{jk}\dot{x}_j\dot{\xi}_k - \Gamma_i^{lb}\dot{x}_l\Gamma_b^{jk}\dot{x}_j\xi_k \tag{16}$$

Equating Eq. (16) and Eq. (13) we obtain

$$\frac{D^2\xi_i}{Dt^2} = \frac{\partial \Gamma_i^{jk}}{\partial x_l}\dot{x}_l\dot{x}_j\xi_k - \frac{\partial \Gamma_i^{jk}}{\partial x_l}\xi_l\dot{x}_j\dot{x}_k + \Gamma_i^{\kappa\nu}\Gamma_\nu^{jk}\dot{x}_\kappa\dot{x}_j\xi_k - \Gamma_i^{bl}\Gamma_b^{jk}\dot{x}_j\dot{x}_k\xi_l +$$

$$+\left(\frac{e}{m}g_{bk}\Gamma_i^{kj}F^{bl}\dot{x}_l - g_{bk}\Gamma_i^{kj}\frac{e}{m}\frac{\partial A^b}{\partial t} - g_{bk}\Gamma_i^{kj}\frac{\partial g^{bl}}{\partial t}\dot{x}_l + \frac{\partial \Gamma_i^{kj}}{\partial t}\dot{x}_k\right)\xi_j + \tag{17}$$

$$+\left(\frac{e}{m}\frac{\partial g_{ib}}{\partial x_k}F^{bj}\dot{x}_j+\frac{e}{m}\frac{\partial F^{bj}}{\partial x_k}g_{ib}\dot{x}_j-\frac{\partial g_{ib}}{\partial x_k}\frac{\partial g^{bj}}{\partial t}\dot{x}_j-g_{ib}\frac{\partial g^{bj}}{\partial x_k \partial t}\dot{x}_j-\frac{e}{m}\frac{\partial A^b}{\partial t}\frac{\partial g_{ib}}{\partial x_k}-\frac{e}{m}g_{ib}\frac{\partial A^b}{\partial x_k \partial t}\right)\xi_k+$$

$$+\left(\frac{e}{m}g_{ib}F^{bk}-g_{ib}\frac{\partial g^{bk}}{\partial t}\right)\dot{\xi}_k$$

Reorganizing terms, we find with the definition

$$R_i^{jkl}=\frac{\partial \Gamma_i^{jl}}{\partial x_k}-\frac{\partial \Gamma_i^{jk}}{\partial x_l}+\Gamma_i^{mk}\Gamma_m^{jl}-\Gamma_i^{ml}\Gamma_m^{jk} \tag{18}$$

of the Riemann curvature tensor, that

$$\frac{D^2 \xi_i}{Dt^2}=R_i^{jkl}\dot{x}_k\dot{x}_j\xi_l+\left(\frac{e}{m}g_{bi}F^{bl}-g_{bi}\frac{\partial g^{bl}}{\partial t}\right)\frac{D\xi_l}{Dt}+$$

$$+\left\{\frac{\partial \Gamma_i^{jl}}{\partial t}-\frac{\partial}{\partial x_l}\left(g_{bi}\left(\frac{e}{m}F^{jb}+\frac{\partial g^{bj}}{\partial t}\right)\right)+\right.$$

$$\left.+\left(g_{bi}\left(\frac{e}{m}F^{kb}+\frac{\partial g^{bk}}{\partial t}\right)\right)\Gamma_k^{jl}-\Gamma_i^{kl}\left(g_{bk}\left(\frac{e}{m}F^{jb}+\frac{\partial g^{bj}}{\partial t}\right)\right)\right\}\dot{x}_j\xi_l+ \tag{19}$$

$$-\frac{e}{m}\frac{\partial A^b}{\partial t}\frac{\partial g_{bi}}{\partial x_l}\xi_l-\frac{e}{m}g_{bi}\frac{\partial A^b}{\partial x_l \partial t}\xi_l-g_{bj}\Gamma_i^{jl}\frac{e}{m}\frac{\partial A^b}{\partial t}\xi_l$$

This equation contains terms additional to the contribution of the curvature tensor **(18)**
In the next section we interpret these additional terms.

**2.3 Interpretation of the geodesic deviation.**

In order to eliminate the first order derivative we redefine the covariant derivative so that the extra argument will cancel out the extra derivative of ξ. First we look at the coefficient of $\frac{D\xi_i}{Dt}$ and define it as:

$$-2\Gamma_i^{4k}=\left(\frac{e}{m}g_{li}F^{lk}-g_{li}\frac{\partial g^{lk}}{\partial t}\right)=-g_{li}\left(\frac{\partial g^{lk}}{\partial t}-\frac{e}{m}F^{lk}\right)=-g_{li}\left(\frac{\partial g^{lk}}{\partial t}+\frac{e}{m}F^{kl}\right)$$

$$=-2\left(\frac{1}{2}g_{li}\left(\frac{\partial g^{lk}}{\partial t}+\frac{e}{m}\frac{\partial A^l}{\partial x_k}-\frac{e}{m}\frac{\partial A^k}{\partial x_l}\right)\right), \tag{20}$$

where we have one anti-symmetric argument and one symmetric as for the form of a connection as in **(3)**, but we shall now write $\frac{\partial g^{ij}}{\partial t}$ as $\frac{\partial g^{ij}}{\partial x_4}$, $\frac{e}{m}\frac{\partial A^i}{\partial x_j}$ as $\frac{\partial g^{4i}}{\partial x_j}$, and identify $\frac{e}{m}A^i$ with a fifth component of the metric, i.e.,

$$g^{4i}=\frac{e}{m}A^i. \tag{21}$$

In this way we shall obtain a result for the Galilean theory with electromagnetism analogy to the structure of the Kaluza-Klein theory. Note that $g^{4i}$ enters formally in the definition of the connection, but $g^{44}$ remains unspecified in the theory. It could be chosen (in an analogy to the choice of Klein [4]) as related to an additional scalar field, corresponding to the addition of a scalar field, to the Hamiltonian **(9)**.

The theory therefore contains the basic structure of the geometry of a 4D Kaluza-Klein theory [5, 4]. As we shall see below, the curvature tensor associated with this manifold also has the form of a 4D Kaluza-Klein type theory.

We may then define

$$\Gamma_i^{j4} = \frac{1}{2} g_{bi} \left( \frac{\partial g^{bj}}{\partial x_4} + \frac{\partial g^{b4}}{\partial x_j} - \frac{\partial g^{4j}}{\partial x_b} \right), \tag{22}$$

and

$$\Gamma_i^{44} = g_{ib} \frac{1}{2} \left( \frac{\partial g^{4b}}{\partial t} + \frac{\partial g^{4b}}{\partial t} - \frac{\partial g^{44}}{\partial x_b} \right) = g_{ib} \frac{\partial g^{4b}}{\partial t} = g_{ib} \frac{e}{m} \frac{\partial A^b}{\partial t}, \tag{23}$$

where we have assumed that $\frac{\partial g^{44}}{\partial x_b} = 0$. This assumption will become clear in the Hamilton space, in which $g^{ij}$ itself contains the "physics" associated with the scalar field $A_0$ (the electric potential). One may think of this as a method for reducing dimensions, as discussed in the conclusions.

Let as rewrite now Eq. **(19)** in terms of the fourth index

$$\frac{D^2 \xi_i}{Dt^2} = R_i^{jkl} \dot{x}_j \dot{x}_k \xi_l + \frac{\partial \Gamma_i^{lj}}{\partial t} \dot{x}_j \xi_l - 2 \frac{\partial \Gamma_i^{4j}}{\partial x_l} \xi_l \dot{x}_j - 2 \Gamma_b^{4j} \Gamma_i^{bl} \dot{x}_j \xi_l$$

$$- \Gamma_i^{lb} \Gamma_b^{44} \xi_l - \frac{\partial \Gamma_i^{44}}{\partial x_l} \xi_l - 2 \Gamma_i^{4l} \dot{\xi}_l \tag{24}$$

Now we have the tools to define a new covariant derivative. Using Eq. **(20)** and **(14)** we can write formally:

$$\frac{\overline{D} \xi_i}{\overline{Dt}} = \frac{d \xi_i}{dt} + \Gamma_i^{bl} \frac{\partial x_b}{\partial t} \xi_l + \Gamma_i^{4l} \frac{\partial t}{\partial t} \xi_l \tag{25}$$

The second derivative now gives

$$\frac{\overline{D}^2 \xi_i}{\overline{Dt}^2} = \frac{D^2 \xi_i}{Dt^2} + 2 \Gamma_i^{4b} \frac{D \xi_b}{D\tau} - \Gamma_i^{4b} \Gamma_b^{jl} \dot{x}_j \xi_l +$$

$$+ \Gamma_i^{jb} \Gamma_b^{4l} \dot{x}_j \xi_l + \Gamma_i^{4b} \Gamma_b^{4l} \xi_l + \frac{\partial \Gamma_i^{4l}}{\partial t} \xi_l + \frac{\partial \Gamma_i^{4l}}{\partial x_j} \dot{x}_j \xi_l, \tag{26}$$

where $\frac{\overline{D}\xi_i}{\overline{Dt}}$ is a new definition of the covariant derivative, and $\frac{D\xi_i}{Dt}$ is the old one.

Substituting equation **(26)** in equation **(20)** and canceling out extra terms in the equation we obtain:

$$\frac{\overline{D^2}\xi_i}{\overline{Dt}^2} = R_i^{kjl}\dot{x}_k\dot{x}_j\xi_l +$$

$$\frac{\partial \Gamma_i^{jt}}{\partial t}\dot{x}_j\xi_l - \frac{\partial \Gamma_i^{4j}}{\partial x_l}\dot{x}_j\xi_l + \Gamma_i^{4m}\Gamma_m^{jl}\dot{x}_j\xi_l - \Gamma_i^{ml}\Gamma_m^{4j}\dot{x}_j\xi_l$$

$$+\frac{\partial \Gamma_i^{4l}}{\partial x_\eta}\dot{x}_j\xi_l - \frac{\partial \Gamma_i^{4j}}{\partial x_l}\dot{x}_j\xi_l + \Gamma_i^{jm}\Gamma_m^{5l}\dot{x}_j\xi_l - \Gamma_i^{ml}\Gamma_m^{4j}\dot{x}_j\xi_l +$$

$$+\frac{\partial \Gamma_i^{4l}}{\partial t}\xi_l - \frac{\partial \Gamma_i^{44}}{\partial x_l}\xi_l + \Gamma_i^{4j}\Gamma_j^{4l}\xi_l - \Gamma_i^{jl}\Gamma_j^{44}\xi_l$$

(27)

where now the first covariant derivative of $\xi$ cancels.

At this point one can make explicit the interpretation of the extra terms we have found as a geometrical consequence of the imbedding. First let us write $\overline{R}_i^{k4l}, \overline{R}_i^{4kl}$ and $\overline{R}_i^{44l}$ by definition in the same way as we defined the generalized connection components (20), i.e.,

$$\overline{R}_i^{4\eta l} = \frac{\partial \Gamma_i^{4l}}{\partial x_k} - \frac{\partial \Gamma_i^{4\eta}}{\partial x_l} + \Gamma_i^{km}\Gamma_m^{4l} - \Gamma_i^{lm}\Gamma_m^{4k}$$

$$\overline{R}_i^{k4l} = \frac{\partial \Gamma_i^{kl}}{\partial t} - \frac{\partial \Gamma_i^{4k}}{\partial x_l} + \Gamma_i^{4m}\Gamma_m^{kl} - \Gamma_i^{lm}\Gamma_m^{4k}$$

(28)

$$\overline{R}_i^{44l} = \frac{\partial \Gamma_i^{l4}}{\partial t} - \frac{\partial \Gamma_i^{44}}{\partial x_l} + \Gamma_i^{4m}\Gamma_m^{l4} - \Gamma_i^{lm}\Gamma_m^{44}$$

The final result for the modified covariant geodesic deviation then takes on the form

$$\frac{\overline{D^2}\xi_i}{\overline{Dt}^2} = \left( R_i^{jkl}\dot{x}_j\dot{x}_k + \overline{R}_i^{4kl}\dot{x}_k\frac{dt}{dt} + \overline{R}_i^{k4l}\dot{x}_k\frac{dt}{dt} + \overline{R}_i^{44l}\frac{dt}{dt}\frac{dt}{dt}\right)\xi_l,$$

(29)

with $\frac{dt}{dt} = 1$.

At this point we can also rewrite the equation of motion Eq. (72) with the new $\Gamma_i^{l4}$ and $\Gamma_i^{44}$, which now appears as a geodesic equation of normal form

$$\ddot{x}_i = -\Gamma_i^{jk}\dot{x}_j\dot{x}_k - 2\Gamma_i^{4k}\dot{x}_k - \Gamma_i^{44},$$

(30)

where we have put all $\frac{dt}{dt}$ terms to unity.

This can be interpreted by assuming a fourth component in g, i.e. $g^{4i}$, and we obtain:

$$g^{4i} = \frac{e}{m}A^i$$

(31)

This leads us to the following form for the equation of motion:

$$\ddot{x}_j = -\Gamma_j^{lm} \dot{x}_l \dot{x}_m - 2\Gamma_j^{4l} \dot{x}_l - \Gamma_j^{44}, \tag{32}$$

where $\Gamma_j^{4l}$ and $\Gamma_j^{44}$ are:

$$\Gamma_j^{4l} = -g_{jk} \frac{1}{2}\left(\frac{\partial g^{lk}}{\partial t} + \frac{\partial g^{4k}}{\partial x_l} - \frac{\partial g^{l4}}{\partial x_k}\right) = -g_{jk} \frac{1}{2}\left(\frac{\partial g^{lk}}{\partial t} + \frac{e}{m}\frac{\partial A^k}{\partial x_l} - \frac{e}{m}\frac{\partial A^l}{\partial x_k}\right)$$

$$= -g_{jk} \frac{1}{2}\left(\frac{\partial g^{lk}}{\partial t} + \frac{e}{m} F^{lk}\right) \tag{33}$$

$$\Gamma_j^{44} = -g_{jk} \frac{1}{2}\left(\frac{\partial g^{4k}}{\partial t} + \frac{\partial g^{4k}}{\partial t} - \frac{\partial g^{44}}{\partial x_k}\right) = -g_{jk} \frac{\partial g^{4k}}{\partial t} = -g_{jk} \frac{e}{m}\frac{\partial A^k}{\partial t},$$

where we have assumed that $\frac{\partial g^{44}}{\partial x_k} = 0$ (corresponding in case $g^{44}$ is the electric potential, to a vanishing electric potential).

The dynamical role of the full Maxwell 4x4 tensor becomes clear under the transformation **(6)**, as we see in the following.

$$\ddot{x}^i = \frac{\partial g^{ij}}{\partial x_k} \dot{x}_k \dot{x}_j + g^{ij} \ddot{x}_j + \frac{\partial g^{ij}}{\partial t} \dot{x}_j \tag{34}$$

Substituting Eq. **(32)** we find:

$$\ddot{x}^i = \frac{\partial g^{ij}}{\partial x_k} \dot{x}_k \dot{x}_j + \frac{\partial g^{ij}}{\partial t} \dot{x}_j - g^{ij} \Gamma_j^{lm} \dot{x}_l \dot{x}_m - \left(\frac{\partial g^{li}}{\partial t} - \frac{e}{m}\frac{\partial A^l}{\partial x_i} + \frac{e}{m}\frac{\partial A^i}{\partial x_l}\right)\dot{x}_l - \frac{e}{m}\frac{\partial A^i}{\partial t} \tag{35}$$

Finally, for the 3D case we obtain:

$$\ddot{x}^i = -\frac{1}{2} g^{im} \frac{\partial g_{jk}}{\partial x^m} \dot{x}^j \dot{x}^k - \frac{e}{m} F_m^{\ i} \dot{x}^m - \frac{e}{m}\frac{\partial A^i}{\partial t} \tag{36}$$

Substituting the form of $g_{ij}$ from **(8)** one then finds

$$\ddot{x}^i = -\frac{1}{2}\frac{E+eA^0}{E}\eta^{im}\frac{\partial}{\partial x^m}\left(\frac{E}{E+eA^0}\eta_{jk}\right)\dot{x}^j \dot{x}^k - \frac{e}{m} F_m^{\ i} \dot{x}^m - \frac{e}{m}\frac{\partial A^i}{\partial t} =$$

$$= \frac{1}{2}\frac{E+eA^0}{E}\eta^{im}\frac{E}{(E+eA^0)^2}\frac{\partial A^0}{\partial x^m}\eta_{jk}\dot{x}^j \dot{x}^k - \frac{e}{m} F_m^{\ i} \dot{x}^m - \frac{e}{m}\frac{\partial A^i}{\partial t} =$$

$$= \frac{1}{2}\eta^{im}\frac{1}{E+eA^0}e\frac{\partial A^0}{\partial x^m}\frac{2}{m}(E+eA^0) - \frac{e}{m} F_m^{\ i} \dot{x}^m - \frac{e}{m}\frac{\partial A^i}{\partial t} = \tag{37}$$

$$= \eta^{im}\frac{e}{m}\frac{\partial A^0}{\partial x^m} - \frac{e}{m} F_m^{\ i} \dot{x}^m - \frac{e}{m}\frac{\partial A^i}{\partial t} = -\frac{e}{m} F_m^{\ i} \dot{x}^m - \frac{e}{m}\left(\frac{\partial A^i}{\partial t} - \eta^{im}\frac{\partial A^0}{\partial x^m}\right)$$

This is the standard Lorentz force in flat space. This result also follows, of course, directly from the Hamilton equation applied to **(7)**. The corresponding geometrical structure in the Gutzwiller space appears to be new, even in this simple example. In this, generalization to the 4D covariant theory, to be treated in the following section, we shall find the well-known Kaluza-Klein structure in the dual space. We now turn to the relativistic problem.

**3. Generalization to 4D Minkowski space**

This method can be extended to motion in space-time using the covariant formulation of Hamiltonian dynamics given by Stueckelberg [6]. In this formulation, the Hamiltonian of the free particle is given by

$$K = \frac{1}{2m}\eta_{\mu\nu} p^\mu p^\nu \tag{38}$$

where $\eta_{\mu\nu} = diag(-1,+1,+1,+1)$ The Hamilton equations are

$$\dot{x}_\mu = \frac{\partial K}{\partial p^\mu}, \quad \dot{p}^\mu = -\frac{\partial K}{\partial x_\mu} \tag{39}$$

where the dot indicates differentiation with respect to the invariant universal parameter of evolution $\tau$. Note that the mass of the particle is, as usual in relativistic dynamics, described by $\eta_{\mu\nu} p^\mu p^\nu = -m_p^2$; the quantity m (taken by Feynman [7] to be unity) is a dimensional fixed parameter associated with the particle.

A model for a system with interaction represented by a potential function V(x) is [6, 8]

$$K = \frac{1}{2m}\eta_{\mu\nu} p^\mu p^\nu + V(x) \tag{40}$$

If we write instead, a Hamiltonian in analogy with the nonrelativistic construction described above, we obtain, with conformal metric [9]

$$g_{\mu\nu} = \eta_{\mu\nu} \frac{K}{K - V(x)} \tag{41}$$

the equivalent form.

$$K = \frac{1}{2m} g_{\mu\nu} p^\mu p^\nu, \tag{42}$$

The Hamilton equations become

$$\dot{x}_\mu = \frac{1}{m} g_{\mu\nu} p^\nu, \quad \dot{p}^\mu = -\frac{1}{2m} p^\nu \frac{\partial g_{\lambda\nu}}{\partial x_\mu} p^\lambda. \tag{43}$$

Again solving for $p^\nu$ from the first of (43), and differentiating, one finds the geodesic equation

$$\ddot{x}_\rho = -\Gamma_\rho^{\mu\nu} \dot{x}_\nu \dot{x}_\mu, \tag{44}$$

where $\Gamma_\rho^{\mu\nu}$ is given by

$$\Gamma_\rho^{\mu\nu} = \frac{1}{2} g_{\rho\lambda} \left( \frac{\partial g^{\lambda\mu}}{\partial x_\nu} + \frac{\partial g^{\lambda\nu}}{\partial x_\mu} - \frac{\partial g^{\mu\nu}}{\partial x_\lambda} \right), \tag{45}$$

which may be recognized as a connection form on the manifold $\{x_\mu\}$.

As a connection form $\Gamma_\rho^{\mu\nu}$ is compatible with $g_{\mu\nu}$. The manifold of solutions of the Hamilton equations is described geometrically in this dual space by the geodesics **(44)**, and the curvature of this manifold is given by the Riemann-Christoffel curvature tensor associated with the connection form **(45)**, i.e.,

$$R_\mu^{\xi\nu\sigma} = \frac{\partial \Gamma_\mu^{\xi\sigma}}{\partial x_\nu} - \frac{\partial \Gamma_\mu^{\xi\nu}}{\partial x_\sigma} + \Gamma_\mu^{\kappa\nu} \Gamma_\kappa^{\xi\sigma} - \Gamma_\mu^{\kappa\sigma} \Gamma_\kappa^{\xi\nu} \tag{46}$$

As for the nonrelativistic case [10], with a mapping of the type **(6)**, **(44)** with $g^{\mu\nu}$, one obtains a geodesic equation of the form

$$\ddot{x}^\mu = -M_{\rho\nu}^\mu \dot{x}^\rho \dot{x}^\nu \tag{47}$$

where

$$M_{\rho\nu}^\mu = \frac{1}{2} g^{\lambda\mu} \frac{\partial g_{\rho\nu}}{\partial x^\lambda}, \tag{48}$$

In the special coordinates in which **(41)** is valid, one obtains the Hamiltonian evolution

$$m\ddot{x}^\mu = -\frac{\partial V}{\partial x_\mu}, \tag{49}$$

In this way, we achieve in **(47)**, a geometrical embedding of the Hamiltonian dynamics.

The effective Hamiltonian for a Stueckelberg type theory with electromagnetic interaction (gauge invariant form) is [11]

$$K = \frac{1}{2m} \eta_{\mu\nu} (p^\mu - ea^\mu)(p^\nu - ea^\nu) - ea_5. \tag{50}$$

where the $\{a_\mu\}$, as fields, may depend on $\tau$ as well as $x^\mu$.

The fifth field $a^5$ is necessary for the gauge invariance of the $\tau$ derivative in the quantum mechanical Stueckelberg-Schrödinger equation

$$\left( i\frac{\partial}{\partial \tau} + ea_5 \right) \Psi_\tau = \frac{1}{2m} \eta_{\mu\nu} (p^\mu - ea^\mu)(p^\nu - ea^\nu) \Psi_\tau. \tag{51}$$

It plays the role of a scalar potential in the Hamiltonian, and can be used to construct a transformation to the general Gutzwiller form, as in **(41)**.

For the general Maxwell equations one finds [11] (by adding a term proportional to $f^{\beta\alpha}f_{\beta\alpha}$ to the action)

$$\partial_\alpha f^{\beta\alpha} = j^\beta, \tag{52}$$

where $j^\beta = (j^\mu, j^5)$, $j^5 = \rho$,

$$j^\mu = -\frac{i}{2m}\left(\psi^*\left(\partial^\mu - iea^\mu\right)\psi - \psi\left(\partial^\mu - iea^\mu\right)\psi^*\right), \tag{53}$$

and

$$f^{\beta\alpha} = \frac{\partial a^\alpha}{\partial x_\beta} - \frac{\partial a^\beta}{\partial x_\alpha}, \tag{54}$$

and $\alpha, \beta = 0,1,2,3,5$.

Note that the fields $a^\mu$ depend, in general, on $(x^\mu, \tau)$. The zero mode of these fields coincides with the usual massless Maxwell fields [11]. To see this, consider the left side of (52),

$$\partial_\alpha \partial^\beta a^\alpha - \partial_\alpha \partial^\alpha a^\beta = \partial^\beta \partial_\alpha a^\alpha - \partial_\alpha \partial^\alpha a^\beta = -\partial_\alpha \partial^\alpha a^\beta \tag{55}$$

where we have used the 5 dimensional analog of the Lorentz gauge, $\partial_\alpha a^\alpha = 0$, so that

$$-\partial_\alpha \partial^\alpha a^\beta = j^\beta \tag{56}$$

This equation is gauge invariant for gauge functions satisfying $\partial_\alpha \partial^\alpha \Lambda = 0$. In particular, $a^5$ satisfies the equation

$$-\partial_\alpha \partial^\alpha a^5 = \rho, \tag{57}$$

where $j^5 = \rho$ is the density of events. Let use the Fourier representation of $a^\beta(x,\tau)$ and $j^\beta(x,\tau)$

$$\begin{aligned} a^\beta(x,\tau) &= \int a^\beta(x,s)\exp(-is\tau)ds \\ j^\beta(x,\tau) &= \int j^\beta(x,s)\exp(-is\tau)ds. \end{aligned} \tag{58}$$

Substituting Eq. (58) in (56)

$$-\partial_\alpha \partial^\alpha \int a^\beta(x,s)\exp(-is\tau)ds = \int j^\beta(x,s)\exp(-is\tau)ds \tag{59}$$

and separating out the fifth derivative term, i.e., derivatives in $\tau$, one obtains

$$\begin{aligned} \left[-\partial_5 \partial^5 - \partial_\mu \partial^\mu\right]\int a^\beta(x,s)\exp(-is\tau)ds &= \int j^\beta(x,s)\exp(-is\tau)ds \\ \left[\sigma s^2 - \partial_\mu \partial^\mu\right]\int a^\beta(x,s)\exp(-is\tau)ds &= \int j^\beta(x,s)\exp(-is\tau)ds \end{aligned} \tag{60}$$

where $\sigma = \pm 1$ depending on the signature of the fifth "coordinate".

It therefore follows that (see also [12])

$$\left[\sigma s^2 - \partial_\mu \partial^\mu\right] a^\beta(x,s) = j^\beta(x,s), \tag{61}$$

One sees that $s^2$ can be considered as the "mass" squared of the off shell field. For an interaction of the form $j^\beta a_\beta$ in the Lagrangian, the mass associated with the wave in the event current $j^\beta$ can be exchanged with the "mass" in the field, as in conservation of four-momentum at a vertex.

It follows from **(52)** that

$$\partial_\beta j^\beta(x) = 0. \tag{62}$$

Let us break this five-divergence up to two parts, i.e., $\partial_\nu j^\nu(x) + \partial_5 j^5(x) = 0$, and integrate over all $\tau$ to obtain

$$\partial_\mu \int j^\mu(x) d\tau = 0. \tag{63}$$

It then follows that

$$J^\mu = \int j^\mu(x) d\tau. \tag{64}$$

is conserved. It can then be identified with the Maxwell current (see also [13]). Then we see that from **(52)**,

$$A^\mu = \int a^\mu(x) d\tau \tag{65}$$

can be identified with the Maxwell fields. They correspond to the zero mode of the $a^\mu$ fields. Since $a^\mu$ and $A^\mu$ differ in dimension, there is a dimensional factor relating the charge $e$ in **(50)** and the (dimensionless) Maxwell charge [11], but we shall not discuss this point further here [12].

Following a procedure similar to that outlined above, with **(50)**, we obtain a geodesic equation which contains the terms appearing in **(44)** but also extra terms in derivatives of $a^\mu$ field. We show that this expression can be understood as a connection form in higher dimension, which coincides with that of the theory of Kaluza and Klein [5, 4].

**4. Geometrical form of the 4D electromagnetic equations**

In this section, we repeat the calculation for the 4D case, for which the Hamiltonian is given by

$$K = \frac{1}{2m} g_{\nu\mu}(p^\mu - ea^\mu)(p^\nu - ea^\nu) \tag{66}$$

where $p^\mu$ is the momentum, $e$ is an elementary charge, $g_{\nu\mu}$ is a metric tensor, and $a^\nu(x,\tau)$ is related to usual Maxwell field as described above, i.e., $A^\nu(x) = \int a^\nu(x,\tau) d\tau$, i.e., the zero mode with respect to the parameter of motion associated with the dynamics induced by the Hamiltonian **(66)**, this generator of evolution describes the dynamics of

Gutzwiller space and may be related to **(50)** by conformal equivalence, as we discuss below in Eqs. **(83)** to **(86)**.

The Hamilton equations derived from **(66)** are:

$$\dot{x}_\mu = \frac{\partial K}{\partial p^\mu} = \frac{1}{m} g_{\nu\mu}\left(p^\nu - ea^\nu\right) \tag{67}$$

$$\dot{p}^\lambda = -\frac{\partial K}{\partial x_\lambda} = -\frac{\partial}{\partial x_\lambda}\left(\frac{1}{m} g_{\nu\mu}\left(p^\mu - ea^\mu\right)\left(p^\nu - ea^\nu\right)\right) \tag{68}$$

From Eq. **(67)**, multiplying by the inverse of $g_{\nu\mu}$, one obtains

$$p^\lambda = g^{\lambda\mu} \dot{x}_\mu + ea^\lambda \tag{69}$$

If we now take the time $(\tau)$ derivative of **(69)**, allowing $g^{\lambda\mu}$ to be $\tau$ dependent, we obtain:

$$\dot{p}^\lambda = \frac{\partial g^{\lambda\mu}}{\partial x_\sigma}\dot{x}_\sigma \dot{x}_\mu + g^{\lambda\mu}\ddot{x}_\mu + \frac{e}{m}\frac{\partial a^\lambda}{\partial x_\sigma}\dot{x}_\sigma + \frac{e}{m}\frac{\partial a^\lambda}{\partial \tau} + \frac{\partial g^{\lambda\mu}}{\partial \tau}\dot{x}_\mu \tag{70}$$

Substituting Eq. **(69)** into Eq. **(68)** we obtain:

$$\dot{p}^\lambda = -\frac{1}{2}\frac{\partial g_{\mu\nu}}{\partial x_\lambda} g^{\mu\xi}\dot{x}_\xi g^{\nu\sigma}\dot{x}_\sigma + g_{\mu\nu}\frac{e}{m}\frac{1}{2}\left(\frac{\partial a^\mu}{\partial x_\lambda} g^{\nu\xi}\dot{x}_\xi + g^{\mu\xi}\dot{x}_\xi \frac{\partial a^\nu}{\partial x_\lambda}\right) \tag{71}$$

Equating Eq. **(71)** and Eq. **(70)** we obtain the equation of motion:

$$\ddot{x}_{\mu'} = -\Gamma^{\xi\sigma}_{\mu'}\dot{x}_\sigma \dot{x}_\xi + \frac{e}{m} g_{\lambda\mu'} f^{\lambda\sigma}\dot{x}_\sigma + g_{\lambda\mu'}\left(-\frac{e}{m}\frac{\partial a^\lambda}{\partial \tau} - \frac{\partial g^{\lambda\mu}}{\partial \tau}\dot{x}_\mu\right), \tag{72}$$

where $f^{\lambda\sigma}$ is the electromagnetic tensor.

This equation contains not only the "geodesic" form (i.e., the first term on the right hand side), described by the Hamiltonian **(66)** as discussed above, but also has contributions due to the presence of the electromagnetic field.

**4.1 Interpretation of the geodesic deviation.**

In this section we introduce the method used in section 2.3 and in a similar way we obtain the following equation for the geodesic deviation using a covariant derivative of the following form

$$\frac{\overline{D}\xi_\mu}{\overline{D}\tau} = \frac{d\xi_\mu}{d\tau} + \Gamma^{\lambda\theta}_\mu \frac{\partial x_\lambda}{\partial \tau}\xi_\theta + \Gamma^{5\theta}_\mu \frac{\partial \tau}{\partial \tau}\xi_\theta. \tag{73}$$

With the definitions

$$\Gamma^{\sigma 5}_\mu = \frac{1}{2} g_{\lambda\mu}\left(\frac{\partial g^{\lambda\sigma}}{\partial \tau} + \frac{\partial g^{\lambda 5}}{\partial x_\sigma} - \frac{\partial g^{5\sigma}}{\partial x_\lambda}\right), \tag{74}$$

$$\Gamma_\mu^{55} = g_{\mu\lambda} \frac{1}{2}\left( \frac{\partial g^{5\lambda}}{\partial \tau} + \frac{\partial g^{5\lambda}}{\partial \tau} - \frac{\partial g^{55}}{\partial x_\lambda} \right) = g_{\mu\lambda} \frac{\partial g^{5\mu}}{\partial \tau} = g_{\mu\lambda} \frac{e}{m} \frac{\partial A^\lambda}{\partial \tau}, \quad (75)$$

Note that $g^{5\sigma}$ enters formally in the definition of the connection, but $g^{55}$ remains unspecified in the theory, as was $g^{44}$ in the previous section. It could be chosen, as we shall do here, to be zero or constant. It could also be chosen, as done by Klein and Gordon ([4]) as related to an additional scalar field. The theory therefore contains the basic structure of the geometry of a Kaluza-Klein theory [5, 4].
As we shall see below, the curvature tensor associated with this manifold is also that of Kaluza-Klein.

Using $x_5 \equiv \tau$ one obtain,

$$\frac{\overline{D^2 \xi_\mu}}{\overline{D\tau^2}} = R_\mu^{\eta\nu\sigma} \dot{x}_\nu \dot{x}_\eta \xi_\sigma +$$

$$\frac{\partial \Gamma_\mu^{\eta\sigma}}{\partial \tau} \dot{x}_\xi \xi_\sigma - \frac{\partial \Gamma_\mu^{5\eta}}{\partial x_\sigma} \dot{x}_\eta \xi_\sigma + \Gamma_\mu^{5\theta} \Gamma_\theta^{\eta\sigma} \dot{x}_\eta \xi_\sigma - \Gamma_\mu^{\theta\sigma} \Gamma_\theta^{5\eta} \dot{x}_\eta \xi_\sigma$$

$$+ \frac{\partial \Gamma_\mu^{5\sigma}}{\partial x_\eta} \dot{x}_\eta \xi_\sigma - \frac{\partial \Gamma_\mu^{5\eta}}{\partial x_\sigma} \dot{x}_\eta \xi_\sigma + \Gamma_\mu^{\eta\theta} \Gamma_\theta^{5\sigma} \dot{x}_\eta \xi_\sigma - \Gamma_\mu^{\theta\sigma} \Gamma_\theta^{5\eta} \dot{x}_\eta \xi_\sigma + \quad (76)$$

$$+ \frac{\partial \Gamma_\mu^{5\sigma}}{\partial \tau} \xi_\sigma - \frac{\partial \Gamma_\mu^{55}}{\partial x_\sigma} \xi_\sigma + \Gamma_\mu^{5\lambda} \Gamma_\lambda^{5\sigma} \xi_\sigma - \Gamma_\mu^{\eta\sigma} \Gamma_\eta^{55} \xi_\sigma$$

Now let us recognize terms for the connection components in a same way as before and obtain (for the non-vanishing components)

$$\overline{R}_\mu^{5\eta\sigma} = \frac{\partial \Gamma_\mu^{5\sigma}}{\partial x_\eta} - \frac{\partial \Gamma_\mu^{5\eta}}{\partial x_\sigma} + \Gamma_\mu^{\eta\theta} \Gamma_\theta^{5\sigma} - \Gamma_\mu^{\sigma\theta} \Gamma_\theta^{5\eta}$$

$$\overline{R}_\mu^{\eta 5\sigma} = \frac{\partial \Gamma_\mu^{\eta\sigma}}{\partial \tau} - \frac{\partial \Gamma_\mu^{5\eta}}{\partial x_\sigma} + \Gamma_\mu^{5\theta} \Gamma_\theta^{\eta\sigma} - \Gamma_\mu^{\sigma\theta} \Gamma_\theta^{5\eta} \quad (77)$$

$$\overline{R}_\mu^{55\sigma} = \frac{\partial \Gamma_\mu^{\sigma 5}}{\partial \tau} - \frac{\partial \Gamma_\mu^{55}}{\partial x_\sigma} + \Gamma_\mu^{5\theta} \Gamma_\theta^{\sigma 5} - \Gamma_\mu^{\sigma\theta} \Gamma_\theta^{55}$$

The following generalize geodesic equation obtain the form

$$\frac{\overline{D^2 \xi_\mu}}{\overline{D\tau^2}} = \left( R_\mu^{\eta\nu\sigma} \dot{x}_\eta \dot{x}_\nu + \overline{R}_\mu^{5\eta\sigma} \dot{x}_\eta \frac{d\tau}{d\tau} + \overline{R}_\mu^{\eta 5\sigma} \dot{x}_\eta \frac{d\tau}{d\tau} + \overline{R}_\mu^{55\sigma} \frac{d\tau}{d\tau} \frac{d\tau}{d\tau} \right) \xi_\sigma, \quad (78)$$

with $\frac{d\tau}{d\tau} = 1$.

### 4.2 Map to the Hamilton space by the locally defined transformation $g_{\mu\lambda}$ in the tangent space and the Lorentz force

As in the non-relativistic case, a transformation of the type **(6)** brings us to the equations of motion that follow from the Hamilton equations applied to **(50)** in the special coordinates for which **(50)** and **(66)** define the conformal metric. By multiplying $\dot{x}_\mu$ by $g^{\rho\mu}$, we define

$$\dot{x}^\rho \equiv g^{\rho\mu}\dot{x}_\mu \tag{79}$$

and taking a derivative with respect to $\tau$, we obtain

$$\ddot{x}^\rho = \frac{\partial g^{\mu\rho}}{\partial x_\eta}\dot{x}_\eta \dot{x}_\mu + g^{\mu\rho}\ddot{x}_\mu + \frac{\partial g^{\rho\mu}}{\partial \tau}\dot{x}_\mu \tag{80}$$

Putting Eq. (30) into Eq. (79) we obtain:

$$\ddot{x}^\rho = \frac{\partial g^{\mu\rho}}{\partial x_\eta}\dot{x}_\eta \dot{x}_\mu + g^{\mu\rho}\left(-\Gamma^{\xi\sigma}_\mu \dot{x}_\sigma \dot{x}_\xi - 2\Gamma^{5\sigma}_{\mu'}\dot{x}_\sigma - \Gamma^{55}_{\mu'}\right) + \frac{\partial g^{\rho\mu}}{\partial \tau}\dot{x} \tag{81}$$

By raising the index of x as in Eq. (79) in order to have consistency throughout the equation we open up the various components of the connection form and we obtain:

$$\ddot{x}^\rho = +\frac{1}{2}g_{\kappa\mu}\frac{\partial g^{\lambda\kappa}}{\partial x_\rho}g_{\lambda\nu}\dot{x}^\mu \dot{x}^\nu - \left(-\frac{e}{m}\partial^\rho a^\mu + \frac{e}{m}\partial^\mu a^\rho\right)\dot{x}_\mu - \frac{e}{m}\frac{\partial a^\rho}{\partial \tau}$$

$$= -\frac{1}{2}g^{\rho\sigma}\frac{\partial g_{\mu\nu}}{\partial x^\sigma}\dot{x}^\mu \dot{x}^\nu - \frac{e}{m}f_\mu{}^\rho \dot{x}^\mu - \frac{e}{m}\frac{\partial a^\rho}{\partial \tau} \tag{82}$$

This is exactly the form of the generalized Lorentz force, associated with (50), as discussed below.

In the conformal case as explained in [6] in which the potential is given by some $a_5$ which is a function of the coordinate alone one can write:

$$K = \frac{1}{2m}\eta_{\mu\nu}(p^\mu - ea^\mu)(p^\nu - ea^\nu) - ea_5 \tag{83}$$

Then comparing with the original geometric Hamiltonian (66), one can obtain an equation for $g_{\mu\nu}$ on the surface K=const (we give condition in an Appendix for K to be constant) which gives:

$$g_{\mu\nu} = \frac{K}{K+ea_5}\eta_{\mu\nu}, \tag{84}$$

By introducing Eq. (84) and (83) into Eq. (82) we obtain:

$$\ddot{x}^\rho = -\frac{1}{2}\frac{K+ea_5}{K}\eta^{\rho\nu}\partial_\nu\left(\frac{K}{K+ea_5}\eta_{\xi\sigma}\right)\dot{x}^\xi \dot{x}^\sigma - \frac{e}{m}f_\mu{}^\rho \dot{x}^\mu - \frac{e}{m}\frac{\partial a^\rho}{\partial \tau} \tag{85}$$

Substituting $\eta_{\xi\sigma}\dot{x}^\xi\dot{x}^\sigma$ with $\frac{2}{m}\left(K+\frac{e}{m}a_5\right)$ we obtain

$$\ddot{x}^\rho = \frac{1}{2}\frac{K+ea_5}{K}\eta^{\rho\nu}\frac{K}{(K+eK_5)^2}e(\partial_\nu a_5)\frac{2}{m}\left(K+\frac{e}{m}a_5\right) - \frac{e}{m}f_\mu{}^\rho \dot{x}^\mu - \frac{e}{m}\frac{\partial a^\rho}{\partial \tau}$$

or,

$$\ddot{x}^\rho = -\frac{e}{m}f_\mu{}^\rho \dot{x}^\mu - \frac{e}{m}\left(\frac{\partial a^\rho}{\partial \tau} - \eta^{\rho\nu}\frac{\partial a^5}{\partial x_\nu}\right) = -\frac{e}{m}f_\mu{}^\rho \dot{x}^\mu - \frac{e}{m}f_5{}^\rho \tag{86}$$

This result is the generalized Lorentz force which follows directly from the Hamilton equations applied to K in the Hamilton form **(50)**.

**Conclusions and Discussion**

We have shown in this work that in the framework of a relativistic dynamics [6, 11] including a 5D electromagnetic field (required for gauge invariance in such a framework), that the Lorentz scalar field associated with this theory can be absorbed into a conformal metric. The geodesics induced by this metric can be cast into the form of a Kaluza-Klein theory.

This procedure is parallel to that worked out by Horwitz et al [14] for the analysis of nonrelativistic Hamiltonian theories. In ref. [10] it was shown that writing the Hamiltonian in terms of a conformal metric, a fully geometrical description of the motion can be developed, governed by the geodesics associated with that metric. The resulting geodesics with **(6)** follow the orbits generated by the original Hamiltonian when the conformal metric is explicitly substituted into the connection form corresponding to the generalized Lorentz force **(86)**.

Passing to the conformal metric, we found that the resulting geodesics can be identified with those of a Kaluza-Klein theory. We have therefore shown that Kaluza-Klein theory can be understood in terms of a conformal transformation of 5D electromagnetism in a flat space, for which the conventional Maxwell fields are the zero mode [6, 11, 15] (see also P. D. Wesson [16]).

This work describes an underlying dynamical structure for Kaluza-Klein type theories, and it may provide a deeper understanding of the structure of higher dimensional gauge theories of this type.

**Appendix: Conservation of K**

In this section, we study the condition that K be a constant of the motion (describing a closed system). The conservation of K is essential in our model, since this forms the structural basis for the conformal metric.

In the Hamilton, space K has the form **(83)**.
With the relation

$$\dot{x}^\mu = \frac{1}{2m}(p^\mu - ea^\mu) \tag{87}$$

we can write

$$K = \frac{m}{2} \dot{x}^\mu \dot{x}^\nu \eta_{\mu\nu} - ea^5. \tag{88}$$

The first term corresponds to the mass squared of the particle, i.e.,

$$\eta_{\mu\nu}(p^\mu - ea^\mu)(p^\nu - ea^\nu) = -(m_p)^2. \tag{89}$$

Therefore

$$\frac{dK}{d\tau} = -\frac{1}{2m}\frac{d}{d\tau}(m_p)^2 - e\frac{da^5}{d\tau}. \tag{90}$$

The total derivative of K then vanishes if

$$e\frac{da^5}{d\tau} = -\frac{1}{2m}\frac{d}{d\tau}(m_p)^2, \tag{91}$$

i.e., the change in mass of the particle is accounted for by a change in the $a^5$ field.

It is clear from Eq. **(57)** that $a^5$ has its source in the mass density (event density), and we see from **(91)** that a closed system of this type contains an exchange of mass between the particle and the field.

In the Gutzwiller form **(66)**,

$$K = \frac{1}{2m} g_{\nu\mu}(p^\mu - ea^\mu)(p^\nu - ea^\nu)$$

or, in terms of $\dot{x}^\mu$,

$$K = \frac{1}{2m} g_{\nu\mu} \dot{x}^\mu \dot{x}^\nu \tag{92}$$

If in this form the resulting connection form is compatible with the metric, there exists coordinate (Fermi-Walker system, or as in general relativity, the local freely falling frame) such that $g_{\nu\mu}$ appears to be locally flat [17, 18].

In this case, locally,

$$K = \frac{1}{2m}\eta_{\nu\mu}\dot{\xi}^\mu\dot{\xi}^\nu, \qquad (93)$$

and since there is no acceleration at this point, $dK/d\tau = 0$. we therefore see the existence of a compatible connection form is closely related to the conservation of mass in the system, i.e., the balance required in **(90)**.

The calculation carried out above had its basis in a conformal transformation from the Hamilton-Stueckelberg form. We find here a condition in the Gutzwiller space that assures the conservation of K. Consider

$$\frac{dK}{d\tau} = \frac{d}{d\tau}\left(\frac{1}{2m}g_{\mu\nu}(p^\mu - ea^\mu)(p^\nu - ea^\nu)\right). \qquad (94)$$

The total derivative of K contains two types of contribution. The implicit dependence due to motions in the phase space vanish due to the Hamilton equations, and therefore only the partial derivative due to the explicit dependence of the fields (and the metric) can contribute. Using the relations (66) and **(94)**, and recalling the definition of $\Gamma_\mu^{5\sigma}, \Gamma_\mu^{55}$, we obtain

$$\frac{dK}{d\tau} = -\frac{m}{2}\frac{\partial g^{\mu\nu}}{\partial \tau}\dot{x}_\mu \dot{x}_\nu - e\frac{\partial a^\mu}{\partial \tau}\dot{x}_\mu = -m\dot{x}_\mu\left(\frac{1}{2}\frac{\partial g^{\mu\nu}}{\partial \tau}\dot{x}_\nu + \frac{e}{m}\frac{\partial a^\mu}{\partial \tau}\right), \qquad (95)$$

and therefore the conservation condition is

$$-\frac{1}{2}\frac{\partial g^{\mu\nu}}{\partial \tau}\dot{x}_\mu\dot{x}_\nu = \frac{e}{m}\frac{\partial a^\mu}{\partial \tau}\dot{x}_\mu \qquad (96)$$

If the total mass is indeed conserved then the relations between the explicit $\tau$ dependence from $g^{\mu\nu}$ and $a^\mu$ to the one in Eq. **(96)**, from which we see that the explicit $\tau$ dependence of the $a^\mu$ fields requires a compensating effect on the geometry.
Note that if $a^\mu$ and $a^5$ have no explicit dependence on $\tau$ that the conservation requirement is automatically satisfied.